\documentclass[a4paper,11pt,twoside]{article}

\usepackage{times}
\usepackage[T1]{fontenc}
\usepackage[english]{babel}
\usepackage{graphicx}
\usepackage{url}
\usepackage{longtable}
\usepackage[final]{listings}
\usepackage[draft]{fixme}
\usepackage[colorlinks, bookmarks]{hyperref}
\usepackage{color}
\usepackage{paralist}
\usepackage{hangcaption}

\usepackage{boxedminipage}
\definecolor{ErrorRed}{rgb}{1,0.80,0.80}
\definecolor{NoteBlue}{rgb}{0.8,0.80,1.0}
\definecolor{correctGreen}{rgb}{0.8,1.0,0.8}

\newcommand{\nop}[1]{}

\newcommand{\snet}{S-Net}

\newcommand{\Fig}[1]{Figure~\ref{#1}}       %ref to figure
\title{CAL: A Language for Aggregating Functional and Extrafunctional Constraints in Streaming Networks%
\thanks{The research leading to these results has received funding
    from the IST FP-7 research project "Asynchronous and Dynamic
    Virtualization through performance ANalysis to support Concurrency
    Engineering (ADVANCE)" under contract no IST-2010-248828.}}

\author
  {
        Alex Shafarenko, Raimund Kirner\\
        \\
        Department of Computer Science\\
        University of Hertfordshire\\
        College Lane, Hatfield AL10 9AB\\
        United Kingdom\\
        \{r.kirner, a.shafarenko\}@herts.ac.uk
  }

\usepackage{listings}
\usepackage{cite}
\usepackage{fancyvrb}
\begin{document}

%%%%%%%%%%%%%%%%%%%%%%%%%%%%%%%%%%%%%%%%%%%%%%%%%%%%%%%%%%%%%%%%%%%%%%%%%%%%
%
% Environment for EBNF style grammar definitions
%
%%%%%%%%%%%%%%%%%%%%%%%%%%%%%%%%%%%%%%%%%%%%%%%%%%%%%%%%%%%%%%%%%%%%%%%%%%%%

\font\euler=cmti12
\newcommand{\sla}[1]{\begin{euler}#1\end{euler}}

\newcommand{\bnfexpr}[1]{{\it #1}}

\newenvironment{BNF}
%{ \vspace*{1ex} \begin{tabular}[t]{p{2.5cm}cl}}
%{ \vspace*{1ex} \begin{tabular}[t]{p{3cm}cl}}
%RK: use \it in environment instead of bHead:
{ \vspace*{1ex} \begin{tabular}[t]{p{3cm}cl}\it}
{ \end{tabular}  }

\newcommand{\bHead}[1]{\textit{#1} & $\Rightarrow$ &}
\newcommand{\bAlt}[2]{\hbox{\sla{(}}\ #1\ $|$\ #2 \hbox{\sla{)}}}
\newcommand{\bOpt}[1]{\hbox{\sla{[}}\,#1\ \hbox{\sla{]}}}
\newcommand{\bStar}[1]{\bOpt{#1}$^{_{\hbox{\normalsize *}}}$}
\newcommand{\bPlus}[1]{\bOpt{#1}$^{_{\hbox{\normalsize +}}}$}
\newcommand{\bOr}{ \\  & $|$ & \ \it }
\newcommand{\bOrL}{\ $|$ \ }
\newcommand{\bSkip}{ \\ &  & \hspace*{2ex}}
\newcommand{\bEnd}{ \\[.8ex] }

\newcommand{\bToken}[1]{~{\textbf{\textrm{\textup{#1}}}}~}
\newcommand{\bSymb}[1]{~{\mbox{\large\texttt{\textup{#1}}}}~}

  % CTCA macros for BNF expressions (snet/docs/latex/sty/bnf.tex)

\maketitle
\thispagestyle{empty}

\begin{abstract}
In this article we present the {\em Constraint Aggregation Language}
(CAL), a declarative language for describing properties of stateless
program components that interact by exchanging messages.
CAL allows one to describe functional as well as extra-functional
behaviours, such as computation latency.
The CAL language intention is to be able to describe the
behaviour of so-called boxes in the context of {\snet}.
However, the language would find application in
other coordination models based on stateless components. 
\end{abstract}

\section{Introduction}
\label{chap:introduction}
%\ADVANCE\ project\bibliography{snet}

The concept of coordination engendered by the coordination language 
{\snet} enforces a strict separation of 
concerns~\cite{GrelSchoShafPPL08,GrelSchoShafPSI06,GrelShafCTCA06}.
The program is represented as a streaming network in which the nodes
are specified as stateless black boxes defined by their interfaces.
An interface definition declares the type of tuples the box agrees to
receive.
It names individual fields without defining the kind of content (e.g.,
numbers, arrays, etc.)  that those are expected to represent.
The interface also defines the output tuples the box is permitted to
produce, in exactly the same manner: as lists of field names.
A box behaves as follows: it is initialised, then an input
tuple is consumed, then output tuples (also called messages), if any, are
produced in some order and then the box terminates.
No input is possible other than via the input tuple and no
state of the box computation is shared with its environment other
than by producing output tuples.

In this report we present the {\em Constraint Aggregation Language}
(CAL), which is a language for specifying constraints about the
program behaviour of boxes.
The CAL interface definition for boxes is presented in the form of 
signature. Here is an example: 
%++++++++++++++++++++++++++++++++++++++++++++++++++++++++++++++++++
\begin{lstlisting}
box boxname ((a,b,n) -> (d,e), (f,g,q,r), ...):
\end{lstlisting}
%++++++++++++++++++++++++++++++++++++++++++++++++++++++++++++++++++
This box takes as input a tuple \verb|(a,b,n)| and
produces tuples of the types \verb|(d,e)|, \verb|(f,g,q,r)|,
\ldots.

Any properties of the objects communicated in and out of a box are
considered dynamic and unknown to the coordinating infrastructure.
In reality, some of them may be static albeit defined in terms of a box language
and thus unavailable to S-Net. The properties are classified into two large categories: 
functional and extrafunctional. Functional properties are concerned with the value of the
object: its shape (e.g., array dimensions), element type, and content.
The extrafunctional properties include, but are not limited to,
computational latency of output tuples, electric power required for
their production, the amount of auxiliary memory that the algorithm needs, etc. 
Generally, extrafunctional properties are those not affecting
the value of the result tuples, but only the efficiency and ``cost'' of their
production.

The purpose of CAL is to enable the programmer to declare
cause-and-effect relations between the input and output
and to define constraints on these.
It is typically the case that functional properties influence the
extrafunctional ones, e.g.: the length $N$ of a 1d array affects the
time required to sort it in a given order and the size of the auxiliary
memory required for the sorting. It is also true that while extrafunctional 
properties are local and tend to be independent from each other, functional 
properties have the tendency to spread out as messages are sent and 
received across a distributed system.

%------------------------------------------------------------------
%------------------------------------------------------------------

\section{The Constraint Aggregation Language (CAL)}

The {\em constraint aggregation language} (CAL) describes properties
of {\snet} and boxes and boxes that share the \snet\ interface concept.
The modular unit of CAL is called a declaration.
A CAL declaration applies to a specific box and is based on the box 
signature.
The top-level grammar of CAL is given in \Fig{fig:bnf-cal-declarations}.
The declaration consists of clauses.
A clause is the statement that the input condition implies the output condition.
The input condition of a box is the conjunction of predicates on terms that include object
variables that refer to the box input.
The output assertion of a box is the conjunction of predicates on terms that include object
variables that refer to the box output. Both kinds of terms may include environment variables
that refer to the state of the environment or the effect of the box on the environment.
The term structure is given in figure \ref{fig:bnf-cal-terms}.

%\verb|Left => Right|

%==================================================================
\begin{figure}
\begin{BNF} %\bOr
\bHead{Declaration}
  Header \bOpt{Decl \bSymb{;}}*
\bEnd

\bHead{Header}
  \bSymb{box} BoxName \bSymb{(} BoxSignature \bSymb{)} \bSymb{:}
\bEnd

\bHead{BoxSignature}
  \bOpt{TupleType} \bSymb{->} \bOpt{TupleType \bOpt{\bSymb{,} TupleType}*}
\bEnd

\bHead{TupleType}
  \bSymb{(} \bOpt{Object \bOpt{\bSymb{,} Object}*} \bSymb{)}
\bEnd

\bHead{Decl}
   \bOpt{Clause} \bOr \bSymb{provided} Conds \bSymb{use} Decl \bSymb{end}
\bEnd

\bHead{Clause}
  Cond \bSymb{=>} Assert % Input-Cond => Output-Cond
\bEnd

\bHead{Cond}
  Predicate \bOpt{\bSymb{,} Predicate}*
\bEnd

\bHead{Assert}
  Predicate \bOpt{\bSymb{,} Predicate}*
\bEnd

\end{BNF}
%  \bnfexpr{PreCond\bSymb{$\sim$}PostCond}
\caption{Top-Level Grammar of CAL\label{fig:bnf-cal-declarations}}
\end{figure}
%==================================================================

%------------------------------------------------------------------

%==================================================================
\begin{figure}
\begin{BNF} %\bOr
 
\bHead{Term}
	Evaluable \bOr General
\bEnd

\bHead{Evaluable}
	Basic \bOr EvTuple \bOr EvHeadTuple
\bEnd

\bHead{Basic}
	Variable \bOr Constant \bOr Symbol 
\bEnd

\bHead{EvTuple}
	 \bSymb{(} Evaluable \bOpt{\bSymb{,} Evaluable}* \bSymb{)}
\bEnd

\bHead{EvHeadTuple}
	 Symbol EvTuple
\bEnd

\bHead{Infix-Exp}
	Prod \bOr add-op Prod \bOr Infix-Exp add-op Prod
\bEnd

\bHead{Prod}
	Basic \bOr Prod times-op Primary \bOr Prod \bSymb{$\,\hat{}\,$} Primary
\bEnd

\bHead{Primary}
	Basic \bOr \bSymb{(} Infix-Exp \bSymb{)}
\bEnd

\bHead{add-op}
	\bSymb{+} \bOr \bSymb{-}
\bEnd

\bHead{times-op}
	\bSymb{*} \bOr \bSymb{/}
\bEnd

\bHead{General}
	Evaluable \bOr Set
\bEnd

\bHead{Set}
	 \bSymb{\{} Evaluable \bOpt{\bSymb{,} Evaluable}* \bSymb{\}} \bOr Variable \bOr Set \bSymb{$\vee$} Set
\bEnd 
\end{BNF}

%  \bnfexpr{PreCond\bSymb{$\sim$}PostCond}
\caption{Term structure in CAL\label{fig:bnf-cal-terms}}
\end{figure}
%==================================================================
%------------------------------------------------------------------

\subsection{Variables and Terms}

A variable is lexically a sequence of letters, underscores and digits preceded by one or two dollar signs, with the first character after the dollar(s) being a letter or undescore. 
A variable can be associated with a {\em value}. There are two types of values: a number and a term, the former being distinguished from the latter lexically. A variable may not be associated with anything; in that case 
it is called a {\em free} variable. All free variables are existentially quantified by CAL. 

Three categories of variables are available to a CAL clause. 
\begin{enumerate}
\item
Object variables. They are lexically identical to the members of the tuple type in the header except that they are preceded by the dollar sign. 
Each of these variables is associated with a term that reflects the properties of the corresponding data object. 
More precisely, the objects on the left-hand side of the signature are assumed to be pre-associated with the property expressions, and those on the righ-hand side will have properties {\em asserted} by the out-condition, but otherwise all object variables behave in the same way. 
\item Environment variables, for example \verb-$$nthreads-, the number of threads available to the box.  They are associated by the environment with values according to the meaning. They are lexically differentiated from the object variables by the double dollar prefix. Note that environment variables have values at run-time, which may change from one activation of
the box to the next, just as object variables.
\item The rest of the variables are {\em local}; they are free and therefore existentially quantified. Their purpose is to facilitate specification of complex conditions.
There exist a special variable  \verb-$_-, which denotes a fresh local variable with a compiler-generated name. It can be associated with a value via unification but cannot be referred to by name to fetch that value, a kind of unification ``black hole''.  
\end{enumerate}

Terms are composed from the basic terms, which are variables, numbers and symbols, using a few constructors. Numbers are rational numbers in the form \verb$n/d$ where both the numerator and the denominator are unsigned integers. \verb$n/1$ can be abbreviated to \verb$n$ and an optional minus may precede any number.

The variety of symbols includes identifiers and infix signs. Identifiers are, as usual, composed of letters, numbers and underscores, and the infix signs include \verb$+, -, *, /, ^, \/$. Infix signs are treated as syntactic sugar for special indentifiers \verb$\plus$, \verb$\minus$, \verb$\times$, \verb$\slash$, \verb$\hat$, \verb$\union$,  which start with a backslash (otherwise not available to the programmer), but which are, otherwise, similar to ordinary identifiers. 

Two constructors are used for term construction: set and tuple. A set is a comma-separated list enclosed in braces, for example: \verb-{$A, 12, shape}-. Sets contain members, which are general terms except any kind of set nesting is forbidden.
Unions of sets are allowed, and a variable may range over sets, even though that is not a type judgement and is not 
enforced by the compiler. Ill-formed sets, which contain sets as members or which are united with members rather than
sets cause the predicate that uses such a set to fail. 

A  tuple is a comma-separated list enclosed in parenthesis. As a syntactic sugar, the first member of a tuple, provided that it is a symbol, can be extracted from it and placed in front of the opening parenthesis. The resulting term is fully equivalent to the original one, for example, \verb$ a(b,c) $ is effectively equivalent to \verb$(a,b,c)$. 

Infix signs implicitly cause tupling. Any term in the form \verb$e1$ $\oplus$ \verb$e2$, where \verb$e1,e2$ are terms and $\oplus$ is an infix sign, is converted by the CAL compiler to $\oplus($\verb$e1$,\verb$e2$ $)$. 
Any ambiguity resulting from more than one infix sign binding the same term is resolved on the basis of priority in a conventional way.

CAL variables have no explicit type, but terms do. Since set nesting is forbidden, there are two mutually incoercible term types: a set and an individual. An individual can be a number, a symbol or a tuple.   

\begin{figure}
\begin{BNF}
\bHead{Predicate}
  \bOpt{Relation \bOrL Equivalence}
\bEnd

\bHead{Relation}
 \bOpt{Variable \bOrL Constant} RelOp Expr
\bEnd

\bHead{RelOp}
\bOpt{
\bSymb{=} \bOrL 
\bSymb{>} \bOrL 
\bSymb{<} \bOrL 
\bSymb{>=} \bOrL 
\bSymb{<=} \bOrL 
\bSymb{!=} 
}
\bEnd

\bHead{Equivalence}
 Expr \bSymb{:=:} Expr
\bEnd

\end{BNF}
\caption{Predicates in CAL}\label{fig:bnf-cal-predicates}
\end{figure}

\subsection{Predicates}

Terms in CAL are used in formulating conditions and assertions, which are conjunctions of {\em predicates}. 
Those have the grammar displayed in Fig \ref{fig:bnf-cal-predicates}. There are two syntactic classes of predicate:
relations and equivalences. 

\paragraph{Relations.} Those compare a variable (or a constant) with a term on the basis of numerical value. 
The term on the right hand side is evaluated in the arithmetic sense as follows
\begin{itemize}
\item Only tuples are allowed; sets are forbidden; if a set constructor is encountered, the predicate fails.
\item The top level of the term structure is thus a basic term or a tuple; 
	If the former,
	\begin{itemize}
		\item if the basic term is a constant, the constant value becomes the value of the term
		\item if the basic term is a symbol, it is one of the Standard Constants (examples are: 
		maximum integer, infinity, etc), otherwise the predicate fails
		\item if the basic term is a variable, the variable must be associated with an evaluable 
		term, otherwise the predicate fails; if it succeeds, then that term's numerical value becomes 
		the value of the variable. 
	\end{itemize}
	If the top level of the term structure is a tuple, the tuple head is interpreted as the identifier 
	of a numerical function, 
	such as \verb$\plus$; the rest of the members of the tuple are considered the arguments 
	and are evaluated; the function is applied to the values of its arguments and the result 
	becomes the value of the term. 
	If the identifier of the function is unknown, the predicate fails, as it does in the event
	of any of the arguments failing.  Note that the tuple head cannot be 
	a variable\footnote{This restriction is required for the tractability of numerical expressions.}.
\end{itemize}
Note that since all free variables are existentially quantified, the clause will effectively define a constraint 
involving every free variable participating in it.

\paragraph{Equivalence} is an application of the most general unifier to the two terms. The unifier succeeds if there exist unifying associations for all free variables participating in the terms.  If it succeeds, the associations will be the most general ones possible, and they may involve further free variables generated by the compiler. The unification rules are completely symmetrical with respect to both operands:

\begin{itemize}
	\item At the top level ground terms, tuples and sets are all possible. 
	Tuples and sets are mutually un-unifiable, unless the tuple head is \verb-\union-. Union-headed tuples are considered sets in the context of unification. It is checked that terms under a union are (syntactically) sets. That excludes, as members of a union, terms such as a tuple headed by anything other than  \verb-\union-, constants and symbols, but it does not exclude variables, as those can, and in this case must, be associated with sets, but that is not a syntactic restriction. 
	\item Sets are unified by assigning the free variables occurring in them with most general terms (i.e. constraining them) that make the two sets identical. By the most general assignment we mean an assignment such that any other unifying assignment can be produced from it by further constraining any free variables that remain after the unification. 
	When sets are unified, the most general form of {\em set term} is, obviously,  \[
	\{t_1,t_2,\ldots,t_n\}\vee v_1 \vee v_2 \vee \ldots \vee v_q
	\,,\] %
	where $t_1,t_2,\ldots,t_n$ are (non-set) terms, i.e. basic terms or tuples of tuples and/or basic terms, and 
	$v_1, v_2, \ldots, v_q$ are variables ranging over sets (since sets and individuals are mutually incoercible)\footnote{
	Unification of such terms is known as the ${\rm\bf flat}(q)$ unification problem\cite{setuni} and it is the most complex set unification problem that admits a polynomial complexity solution. It is known that even the solution to a system of  ${\rm\bf flat}(q)$ unification equations, as opposed to a single equation, is already NP-complete, to say nothing of a more complex set algebra. This explains our choice of set operators available in CAL.}
	\item A non-variable basic term can only be unified with an identical basic term or a variable.
	\item An un-associated variable becomes associated with the other operand; if the variable is associated, the other operand is unified with the associated terms.
	  
\end{itemize}

\subsection{A note on logic/constraint programming machinery}

It is easy to see that the present syntax of CAL clauses makes them reducible to Horn clauses. Indeed, 
due to the static nature of S-Net routing, the assertions can be matched with conditions using the type-inferred 
version of the S-Net graph. As a result, after a suitable (fully mechanical) transformation, a single unstructured 
set of CAL clauses can be gleaned from the whole network. 
The only deviation from the standard Horn format:\[
P_1 \vee \bar{P_2} \vee \bar{P_3}\ldots   
\]
is the conjunctive head made up from all assertions $A_1$, $A_2$, etc. in the clause
\[
(A_1 \wedge A_2 \wedge A_3 \ldots) \vee \bar{C_1} \vee \bar{C_2}\ldots \,,
\]
with the conditions $C_1$, $C_2$, etc. participating in the disjunction.
However, this is equivalent to a set of Horn clauses:\[
\begin{array}{ccc}
&A_1 \vee \bar{P_1} \vee \bar{P_2}\ldots &\nonumber \\
&A_2 \vee \bar{P_1} \vee \bar{P_2}\ldots & \nonumber \\
&A_3 \vee \bar{P_1} \vee \bar{P_2}\ldots  &\nonumber \\
&\ldots& \nonumber 
\end{array} 
\]
The central issue therefore is whether or not the assertions require a disjunctive form, which would make CAL 
clauses non-Horn. The present syntax excludes it in anticipation that the input information should be sufficient to assert constraints (however loosely) on a definite collection of terms, each member of the collection thus being 
constrained conjunctively. It may be the case, however, that an input to a box produces, for instance, two possible 
output messages; then depending on which message is produced, a different constraint is asserted on the same object variable. Such situations would necessitate a disjunctive assertion, for which the current CAL syntax is insufficient. 

Thoughts should be given to non-Horn CSPs, in our case SMT modulo linear arithmetic, which would involve
a solver such as Yices\cite{yices} as a means of aggregation.       

\subsection{Aggregation and Vocabularies}

The idea behind constraint specification of CAL is that both functional and extrafunctional constraints can be expressed
as relations on terms.  The difference between them is in the aggregation method. For
functional constraints terms are unified according to the corresponding box connections. 
For instance, two boxes connected in series:
\verb-A..B- with the signatures, respectively, \verb$(x)->(y)$ and \verb$(p)->(q)$ would impel CAL to unify object variables $y$ and $p$, which will effectively transform any assertions on $y$ into conditions on $p$.
All kinds of connections between boxes will be type-resolved by the coordination language compiler to establish
the interface type; then CAL will unify corresponding object variables. Naturally, due to the presence of arithmetic relations in CAL clauses, the connection between the input and the output of a box might be quite complex and the corresponding aggregation might involve heuristically driven symbolic computations, but such scenarios are expected to be uncommon. At any rate, if constraints fail to aggregate, this leaves the whole-system constraint only set less tight, but still sound. Warning messages could be generated for the benefit of the system designer in such a case, and improvements could include redesign of some boxes in order to expose essential parameters in a more straightforward fashion {\em in data} before any properties and property associations may be formulated. For instance, instead of defining a complex relation between input and output shapes, a box could compute an extra integer parameter that defines the output shapes directly. The output constraints would thus become disconnected from the input constraints, but the former could then be declared precisely and simply. It is clear from this example that there may well be a trade-off between aggregability and accuracy given a limited amount of deductive power in the constraint aggregation logic.

In contrast to functional constraint aggregation, the extrafunctional constraints are aggregated by means other than unification. Unification can still play a role if aggregation rules are formulated in inference-rule form, but that is an  implementation issue. What is certain in any implementation is that the aggregate constraint set is produced from the assertions of individual boxes by non-logical means.  For example, aggregating latency over a pipeline involves not only the knowledge of latency properties of the pipeline stages, but also the communication cost of the pipeline itself, and all those are combined statistically as probability distribution functios with tuneable parameters. The machinery that should be used here is one of queuing theory; also involved in these calculations are the virtual hardware model and various system parameters. At this point it is envisaged that any component-wise information about latency is fed into a special aggregation library that contains an aggregating function for every network combinator per extrafunctional property to be aggregated. It is important to understand that here virtual hardware plays a dual role: it provides 
an execution platform and also
some symbolic rules that describe the process of statistical aggregation relevant to that platform. For example, the above mentioned pipeline example \verb-A..B- would exhibit different temporal behaviours (even statistically) if the nature
of the serial combinator is network communication between processors rather than shared memory for cores. This difference may be qualitative rather than quantitative: in the former case the message size negatively impacts on the combinator latency while in the latter it may not. By contrast, the combinator jitter may behave in the opposite manner: communication tends to reduce jitter due to buffering while direct sharing exacerbates it by allowing the box jitters to combine. 

In both functional and extrafunctional cases the recursive structure of the network would call for a reflexive aggregation technique: subnetworks should expose the same nature collections of symbolic expressions as boxes, and the aggregation process should not care whether what is being aggregated is atomic or not. There are two ways of achieving this. One is to see CAL as a package operating on top of a generic inference system, such as Prolog, or even an appropriate SMT solver such as Yices. This would enable the development of all aggregation code in that framework. The second alternative is to use CAL specifications as source data to a custom analysis tool, which is hardwired for all forms of aggregation. The choice remains open at present time. 

Either form of aggregation requires a definition of the property terms with which the aggregation process must deal. 
CAL terms are generally tuples or sets of tuples, and those involve constants and identifiers. The latter must carry meaning for each individual property class. The structure of the terms and the assortment and meaning of the identifiers used in them form the substance of a {\em vocabulary definition} for a property. Vocabularies for extrafunctional properties are external to CAL and should be defined as part of the contract between the statistical model, coordination infrastructure and virtual hardware. The last one has its own vocabulary that defines the system parameters repeatedly mentioned above. Vocabularies for functional properties are also external to CAL, and are part of a contract, too. However, the contract is now between modules written in a box language. 

Constraint aggregation may deliver information to each of the modules that standard type systems are unable to guarantee and, because of that, fail to gather. CAL properties can be used to tune the back end of the compiler for program specialisation, either statically or using just-in-time compilation techniques, if the CAL process is use dynamically, triggered by a change of value for an environment variable. This would enable quasistatic properties, i.e. properties that are dynamic but which change infrequently, to be treated as configuration parameters, and the compilation process as one of reconfiguration under the influence of nonlocal information.  In contrast to types, 
failure to gather the necessary information would only delay or cancel such reconfiguration but not jeopardise the validity of code, hence more liberal aggregation methods, including heuristic and constraint-satisfaction modulo generally undecidable theories could be tried. Vocabulary-wise, the vocabulary for the functional properties will closely reflect (but not necessarily coincide with, up to isomorphism) the type system of the box language. The clauses would analyse the properties and synthesise assertions that link up types. 

Importantly, types can be dependent on integers communicated at run time, since those can be aggregated and constrained by the CAL inference machinery. As a side effect of the aggregation, certain environment variables may receive assertions, and those can be used by the box-language compiler in its conditional compilation facilities, such as C's \verb-#define-. This would require a convention on such environment-variables' names, for instance they could start with \verb-$$_-. It would also require a template for the \verb-#define--like statements for the box language into which the variables will be substituted before box (re-)compilation.  Interestingly, this approach erodes the boundary between the static and the dynamic in box compilation. 

Below we will present some example vocabularies. The examples are meant to be illustrative rather than definitive and are not prescriptive in any way.

\subsubsection{Functional Vocabulary}

It would be sensible to ensure that any object variable is associated with a set of terms. This way specific properties
can be expressed as one or more terms and the input condition can extract specific terms by set unification, while 
not caring about what other term structures may be present in the set. 
 
The functional vocabulary in a simple Fortran-like programming language with arrays may include the following term:
{\small
\begin{Verbatim}[frame=single,rulecolor=\color{blue},fontfamily=helvetica]

Type(array, element($eltype), rank($r), 
        shape($s0,($s1,($s2,nil))) )

\end{Verbatim}  

}
Here the term head \verb$Type$ indicates that the term defines the object's type. 
Symbol \verb$array$ indicates that the type is an array type. The \verb-element- subterm defines 
the element type of the array as \verb-$eltype-, the rank subterm defines the rank of the array as \verb-$r-
and the shape list is contained in the term \verb-shape-. Note that the element type can be a symbol
that serves as a reference to another type declaration. For example, the following pair of terms define 
a vector of vectors of integers, the inner vector being of variable length:

\vbox{{\small
\begin{Verbatim}[frame=single,rulecolor=\color{blue},fontfamily=helvetica]

Type(this, element(vector), rank(1), 
       shape(100,nil) )

type(vector, element(int), rank(1), 
       shape(unknown, nil) )


\end{Verbatim}  
}}
The second term uses the tuple head \verb-type- rather than \verb-Type- to indicate an additional type definition.
Also note the symbol for unknown integer number \verb$unknown$. The action of arithmetic operators in relational contexts is extended to include the standard response to unknown numbers\footnote{details to follow}.

It should be clear by now that, for instance, all non-pointer C types can be encoded as CAL terms. Records and unions
can be dealt with by using tuples of label-type pairs, both labels and types represented by CAL symbols. Self-referencing
types are possible by reusing the type name symbol (second member of the type tuple at the top level) in the component
terms of the type. Here is a list of integers:
{\small
\begin{Verbatim}[frame=single,rulecolor=\color{blue},fontfamily=helvetica]

Type(int_list, union( record((head, int),tail(int_list)), nil ) )

\end{Verbatim}  
}

Another kind of functional constraint that CAL may deal with is a constraint associated with a value. Values of compound
objects are box-language dependent and consequently intractable in CAL, but integer scalars should be quite tractable
and do usually carry important information that affects functional properties of objects (such as their dimensions).
To this end, the term \verb-value($v)-
should be included in the vocabulary, where \verb-$v- is an integer constant. 
Here is an example of full terms associated with an input interface of \verb$(A,K)->$...
{\small
\begin{Verbatim}[frame=single,rulecolor=\color{blue},fontfamily=helvetica]

$A :=:  {Type(array, element(real), rank(2), shape(7,(7,nil)) ), packed(row_major)}
$K :=:  {Type(int), value(100)} 

\end{Verbatim}  
}
Here the first line associates a term that declares a real $7\times7$ array packed in a row major order, and the second line 
a term that declares the integer scalar $100$. 

\subsubsection{Extrafunctional Vocabulary: Latency}

The latency vocabulary describes the statistical model of message production by a box. The environment provides
a series of variables in its vocabulary: \verb-$$T0-, \verb-$$T1-, \verb-$$T2-, etc. These variables are provided
separately for each box, just as all other double-dollared variables. They are part of the environment input;
consequently they are unassociated. Their number corresponds to the number of type alternatives on the right
hand side of the box type signature. For every output type $\tau_n$, the variable \verb-$$T-$n$ should be unified
with the time-complexity model of the corresponding output, for example, the assertion
{\small
\begin{Verbatim}[frame=single,rulecolor=\color{blue},fontfamily=helvetica]

$$T0 :=: $N * log($N)

\end{Verbatim}  
} will tell the environment that the time complexity of producing one output of the first output type is $N\log N$,
where $N$ is a local variable (encoded as \verb-$N-) constrained somehow by the input condition. $N$ could, for instance,
be unified with one of the object's array dimensions, or the value of a scalar integer supplied through the input 
interface of a box. It is also possible that $N$ has an arithmetic connection with the input data, such as $N=M+1$,
where $M$ comes from the input object set\footnote{This is also true of functional constraints, they may involve
arithmetic relations, which is why CAL introduces the machinery of rational numbers. The hope at this point is that
for most practical purposes linear arithmetic is sufficient.}.

The complexity language used for defining latency models should include symbols for multiplication, \verb-\times-,
exponentiation \verb-\hat-, division \verb-\slash- and logarithm \verb-log-. 

There can be more than one form of complexity. For instance, the latency of a type channel could be probabilistic:
the number of messages produced within time $t$ is described by the Poisson distribution $P_t(\lambda)$ with 
some parameter $\lambda$. This, for example, could be represented by   
{\small
\begin{Verbatim}[frame=single,rulecolor=\color{blue},fontfamily=helvetica]

$$T0 :=: Poisson(1/$N)

\end{Verbatim}  
} where $1/N$ is the $\lambda$-parameter and $N$, as before, an input-related scalar. Naturally, an assertion such as 
{\small
\begin{Verbatim}[frame=single,rulecolor=\color{blue},fontfamily=helvetica]

$$T0 :=: Poisson(unknown)

\end{Verbatim}  
} should also be possible.

Finally, it should be said that the number of messages produced by the box into each of the output type channels
should be constrained whenever possible and the constraints should be communicated to CAL. That can be done in a way
similar to the latency constraints, by employing environment variables \verb-$$M0-, \verb-$$M1-, \verb-$$M2-, etc. 
They should be associated with either integer numbers that indicate how many messages of each type will be produced, or a
term such as 
{\small
\begin{Verbatim}[frame=single,rulecolor=\color{blue},fontfamily=helvetica]

$$M0 :=: limits(5,15)

\end{Verbatim}  
}
indicating the limits of variation, or indeed could be unknown or unbounded (either taken as the default
in the absence of an assertion). 

Here are some examples illustrating the use of CAL
{\small
\begin{Verbatim}[frame=single,rulecolor=\color{blue},fontfamily=helvetica]
box MYBOX: (a,k) => (b), (c,d)

provided $a :=: 
     {Type(array, element($t), rank(2), shape($n,($m,nil)) )} \/ $_ , 
	       $k:=: {value($kv), Type(int)} \/ $_
use
	=>  $n1=$n+1, 
	      $base :=: 
	      {Type(array, element($t), shape ($n1,($m,nil)) )};	        -- Clause 1
	
	=> $b :=: $base \/ {rank(2)}, 
	     $d = $base \/ {rank(3)};							-- Clause 2
	
	$kv > $$nthreads * 100
		=>
	$$T0 :=: $m * log($m)/$$nthreads, $$T1 :=: 1;			-- Clause 3
	
	$kv <= $$nthreads* 100
		=> 
	$$T1 :=: $m^(3/2); $$M1 :=: 0;						-- Clause 4
end 
\end{Verbatim}

}
Four clauses are declared here, all input-dependent on the {\tt\bf provided} condition. The latter unifies the object variables from the input tuple with sets of terms, thus associating local variables. The object variable \verb-$a- is examined to effectively determine whether the corresponding object has rank 2, shape $n\times m$ and element type $t$.
If any part of this information  is not available, no assertion will be made by MYBOX.

Next, Clauses 1 and 2 have null input conditions. They assert the ranks and shapes of output objects $b$ and $d$.
Object $b$ is a two-dimensional object $(n+1)\times m$ of the same type as $a$. Object $d$ is three-dimensional, 
but the third dimension is not constrained by MYBOX a priori. 

Clauses 3 and 4 constrain the environment variables \verb-$$T-$n$.  
For the first tuple type, the result will be computed in $O(m\log{m}/n_{\rm th})$ time, and for the second output tuple the expected computational time is $O(1)$. Both constraints are predicated on a sufficiently large value of the integer scalar
parameter $k$: $k> 100 n_{\rm th}$. The environment variable \verb-$$nthreads- is associated with the value $n_{\rm th}$, which is the number of independent hardware threads available to the box in question.   When $k\le 100 n_{\rm th}$,
the expected computation time changes (this could be due to the box using a different algorithm): now it is, for the first tuple type, $O(m^{3/2})$ and it does not depend on the number of threads (for instance, because the box uses a sequential algorithm for small object sizes). Under such conditions, the second message type is not produced at all, which 
is indicated by constraining the environment variable \verb-$$M1- to the value 0.

\section{Conclusion} 
\label{chap:conclusion}

This article describes the basic concepts of the 
{\em Constraint Aggregation Language} (CAL), a novel program-behaviour
description language to be used to describe the functional and
extra-functional behaviour of software components as terms
in a term algebra, to serve as data for a constraint solver.
While CAL may evolve over time to extend its expressiveness to further
patterns of program behaviour, the initial version presented in this
article already shows the core language features of CAL and the basic
declaration mechanisms provied by CAL.

\bibliography{literature}{}
\bibliographystyle{plain}

\end{document}